\documentclass[twocolumn,showpacs,prb,footinbib,a4paper,superscriptaddress,floatfix]{revtex4}

\usepackage{amssymb,amsmath}
\usepackage{graphics}
\usepackage{dcolumn}
\usepackage{bm}
\usepackage{epsfig}
\usepackage[usenames,dvipsnames]{color}

\begin{document}

\title{Nonequilibrium distribution functions for quantum transport: \\
universality and approximation for the steady state regime}

\author{H. Ness}
\altaffiliation{Present address: Department of Physics, School of Natural and Mathematical Sciences,
King's College London, Strand, London WC2R 2LS, UK}
\email{herve.ness@kcl.ac.uk}
\affiliation{Department of Physics, University of York, Heslington, York YO10 5DD,UK}
\affiliation{European Theoretical Spectroscopy Facility (ETSF)}

\date{\today}

\begin{abstract}
We derive a general expression for the electron nonequilibrium (NE) distribution function
in the context of steady state quantum transport through a two-terminal nanodevice with interaction.
The central idea for the use of NE distributions for open quantum systems is that
both the NE and many-body (MB) effects are taken into account in the statistics of the 
finite size system connected to reservoirs. 
We develop an alternative scheme to calculate the NE steady state properties 
of such systems. The method, using NE distribution and spectral functions,
presents several advantages, and is equivalent to conventional steady-state
NE Green's functions (NEGF) calculations when the same level of approximation for the MB 
interaction is used.
The advantages of our method resides in the fact that the NE distribution and spectral 
functions have better analytic behaviour for numerical calculations.
Furthermore our approach offer the possibility of introducing further approximations, 
not only at the level of the MB interaction as in NEGF, but also at the level of 
the functional form used for the NE distributions.
For the single level model with electron-phonon coupling we have considered, such approximations 
provide a good representation of the exact results, for either the NE distributions themselves 
or the transport properties.
We also derive the formal extensions of our method for systems consisting of 
several electronic levels and several vibration modes.
\end{abstract}
 

\maketitle

\section{Introduction}
\label{sec:intro}

The understanding of irreversible phenomena including nonequilibrium (NE) steady state 
is a long-standing problem of quantum statistical mechanics. 
With the recent experimental developments, it is now possible to measure the transport 
properties through nanoscale systems. These can be either the electronic charge transport
or heat transport. Both properties, i.e. thermoelectric transport, have recently been 
measured simultaneously \cite{Widawsky:2012}.
Such properties exhibit many important new features in comparison with conduction
through macroscopic systems. In particular, the interactions, such as Coulomb interaction 
between electrons and scattering from atomic vibrations, become critically important
in nanoscale objects, especially in single organic molecules \cite{DiVentra:book2008,Cuevas:book2010}

Modelling such transport properties is still a challenge since one needs to be able to
describe the system at the atomic level in a realistic manner, and one needs to use
a formalism for the quantum transport that takes full account of the NE conditions
(full nonlinear response) and the many-body (MB) interaction.

Nonequilibrium Green's functions (NEGF) seems, at the present moment, the best way to
tackle the problem. However, NEGF calculations for realistic systems are difficult to
achieve, beyond mean-field-like approximations or quasi-equilibrium regime, since
the calculations of the MB effects for a large number of electronic (and vibronic)
degrees of freedom are extremely demanding.
Alternatively, the density-functional (DF)-based theories can handle large systems, but 
unfortunately treat the interaction (between electrons for example) on a mean-field-like
basis and the corresponding functionals are not necessarily optimized, or even valid,
for the NE conditions.

In this paper, we present an alternative approach based on the use of NE distribution
and spectral functions. On one hand, such an approach is, in principle, strictly equivalent
to the steady-state NEGF technique, since there is a one-to-one equivalence between the Green's
functions (GF) and the NE distribution and spectral functions. One the other hand, approximations
for the MB effects (in the presence of NE conditions) seems to be more easily
introduced in the NE distribution, while keeping a clear physical interpretation. Furthermore
the use of approximated NE distributions may offer an alternative approach for future
implementations in DF-based calculations for large systems.

In earlier studies, we have already started developing and using the concept of NE distribution 
functions. This was done in a critical analysis of the applicability of Landauer formalism for 
NE current in the presence of interactions \cite{Ness:2010}, and in the study of 
the NE charge susceptibility and its relation with the nonlinear dynamical 
conductance \cite{Ness:2012}.

In this paper, we develop in detail our approach using NE distribution and 
spectral functions, and provide numerical applications.
The paper is organized as follows.
In Sec. \ref{sec:NESS}, we define the general steady state transport set-up. We
start by considering a model system in Sec. \ref{sec:onelevel} and provide
all the analytical results for the NE distributions.
Sec. \ref{sec:fNE} concerns the general properties of the NE distributions.
In Sec. \ref{sec:fNE_algo}, we develop an algorithm for performing NE calculations.
Numerical applications are provided in Sec. \ref{sec:numerics} where we show examples
of the NE distributions for a model of electron-phonon interacting system. The performance
of the exact and approximated NE distributions are studied in this Section.
The generalisation of our approach to more realistic systems are provided in Sec. \ref{sec:realistic}.
Finally we comment our results and conclude our study in Sec. \ref{sec:ccl}.

\section{Steady state quantum transport}
\label{sec:NESS}

We consider a system consisting of a central region $C$ connected to 
two non-interacting Fermi seas. 
The left ($L$) and right ($R$) electrodes are at their own equilibrium, with a Fermi 
distribution $f_\alpha(\omega)$ defined by their respective 
chemical potentials $\mu_\alpha$ and temperatures $T_\alpha$ ($\alpha=L,R$).
The central region $C$ connected to the leads contain interaction
characterized by a self-energy $\Sigma_{\rm int}(\omega)$ in the NEGF formalism.
Furthermore the specific model used for the leads does not need to be specified
at the moment, as long as the leads can also be described by an embedding
self-energy $\Sigma_{\alpha}(\omega)$ in the electron GF of the central region.

The possibility of reaching a steady state regime in such a two-terminal
device has been explored by many authors.
The full time-dependent NEGF formalism and the influence of bound states in
the central region have been studied in 
Refs.~[\onlinecite{Stefanucci:2004a,Stefanucci:2007,Myohanen:2008,Myohanen:2009,Stefanucci:2013}].
Rigorous mathematical methods based on the $C^*$ algebra have been used to
study the existence and stability of such NE steady state, i.e., its independence 
of the way the division into subsystems and reservoirs is performed and its stability
against local perturbations, in the absence \cite{Ruelle:2000,Tasaki:2003,Tasaki:2006,Tasaki:2011}
and in the presence of interaction \cite{Moldoveanu:2011,Cornean:2013}.

For an established steady state regime, it is expected that some formal advantages 
may be given by an approach to NE processes in which the Gibbs-like ensembles play
a prominent role. The construction of such Gibbs-like ensembles for the NE steady 
state can be obtained either by using the MacLennan-Zubarev approaches 
\cite{McLennan:1959,Zubarev:1974,Zubarev:1994,Zubarev:1996,Zubarev:1997,Morozov:1998,Tasaki:2003,Maes:2010}
or the NE density matrix approach developed by Hershfield in Ref.~[\onlinecite{Hershfield:1993}].
The latter has been extensively used for calculating quantum electron transport properties, with or
without interaction \cite{Schiller:1995,Schiller:1998,Han:2006,
Han:2007,Han:2007b,Han:2010,Han:2010b,Dutt:2011,Han:2012}.

In the following, we show that the NE statistics of the open quantum system, i.e. the central 
region $C$ contains, information not only of the NE conditions but also about the MB interaction.

\section{The single-impurity model}
\label{sec:onelevel}

We now consider a model for the central region made of a single electron level in the presence
of interaction. In this section all quantities are either real or complex number functions of
a single energy argument.

\subsection{The NE distribution $f^{\rm NE}$}
\label{sec:fNE}

In a recent paper \cite{Ness:2013}, we have shown, using MacLennan-Zubarev and Hershfield approaches,
that the steady state can be interpreted as an effective equilibrium state with
a corresponding NE density matrix, or equivalently, with a corresponding NE statistics.

Such a NE statistic can be defined by a NE distribution function $f^{\rm NE}(\omega)$. It
enters the relation between the different GFs defined in the central region $C$ as follows: 
\begin{equation}
\label{eq:GF_and_fNE}
G_C^\lessgtr(\omega)  = -  f^{\rm NE,\lessgtr}(\omega)    \left( G_C^r(\omega) - G_C^a(\omega) \right) \ ,
\end{equation}
with $f^{\rm NE,<}(\omega)=f^{\rm NE}$ and
with $f^{\rm NE,>}(\omega)=f^{\rm NE}-1$ .
We recall that the spectral function $A_C(\omega)$ of the central region is obtained
from  $A_C(\omega) = ( G_C^a - G_C^r) / i 2\pi$.
Eq.~(\ref{eq:GF_and_fNE}) bears resemblance with the so-called Kadanoff-Baym Ansatz
\cite{Kadanoff:1962,Lipavski:1986}, but as we have shown in Ref.[\onlinecite{Ness:2013}], 
it is a strictly exact result for the steady state regime.

At equilibrium, $f^{\rm NE}$ is simply the Fermi distribution $f^{\rm eq}$. Out of equilibrium,
the distribution function will depend on the set-up, i.e. on the forces driving the system
(gradient of chemical potential and/or temperature between the leads), and on the interaction
present in the region $C$.

In the absence of interaction, the NE distribution
function for the electron is simply given by \cite{Hershfield:1991,Schiller:1998}
\begin{equation}
\label{eq:f0NE}
f_0^{\rm NE}(\omega)=\frac{\Gamma_L(\omega)f_L(\omega)+\Gamma_R(\omega) f_R(\omega)}
{\Gamma_L(\omega)+\Gamma_R(\omega)}
\end{equation}
where $\Gamma_\alpha(\omega)=i(\Sigma_\alpha^>-\Sigma_\alpha^<)(\omega)$ is the spectral function
of the embedding (lead $\alpha$) self-energy.
It is simply a double-step function, with more or less steep
steps (depending on the temperature $T_L$ and $T_R$) located around $\omega=\mu_L$ and $\omega=\mu_R$, 
and separated by $\mu_L-\mu_R=eV$ ($V$ is the applied bias).

In the presence of interaction in the central region $C$, the NE distribution is given by \cite{Ness:2013}
\begin{equation}
\label{eq:fNE_a}
\begin{split}
f^{\rm NE}(\omega) & = \frac{G_C^<}{G_C^a - G_C^r}
=\frac{G^r_C \Sigma^< G^a_C}{G^r_C \left( (G^r_C)^{-1}-(G^a_C)^{-1}\right) G^a_C} \\
& =\frac{\Sigma^<_L+\Sigma^<_R+\Sigma^<_{\rm int}}{\Sigma^a-\Sigma^r} .
\end{split}
\end{equation}
Using the definitions $\Sigma^<_L+\Sigma^<_R=i \Gamma_{L+R}f_0^{\rm NE}$
and $\Sigma^a-\Sigma^r=-(\Sigma^>-\Sigma^<)=i\Gamma_{L+R}-(\Sigma^>_{\rm int}-\Sigma^<_{\rm int} )$,
with $\Gamma_{L+R}(\omega)=\Gamma_L(\omega)+\Gamma_R(\omega)$,
we obtain:
\begin{equation}
\label{eq:fNE}
f^{\rm NE}(\omega) = \frac{f_0^{\rm NE}(\omega) - i \Sigma^<_{\rm int}(\omega)/\Gamma_{L+R}(\omega)}
{1+i (\Sigma^>_{\rm int}-\Sigma^<_{\rm int} ) / \Gamma_{L+R} } \ .
\end{equation}

Eq.~(\ref{eq:fNE}) is the ``universal'' expression of the electron NE distribution function. 
It is universal with respect to the interaction, in the same sense 
that the GFs have an universal expression via the use of the interaction self-energies.
However, as expected for NE conditions, the NE distribution function is not as 
universal as its equilibrium counterpart, since it depends on both the set-up that drives the system
out of equilibrium (via $f_0^{\rm NE}$) and on the MB interaction $\Sigma_{\rm int}^\lessgtr$ 
(which are themselves dependent on the NE conditions).
We comment more on these properties in Appendix \ref{app:fNE_withwithout_inter}.

From Eq.~(\ref{eq:fNE}), we can see that the NE distribution $f^{\rm NE}$ arises from
two terms 
\begin{equation}
\label{eq:fNE_decomp}
f^{\rm NE}(\omega)=\tilde f_0^{\rm NE}(\omega)+\delta f^{\rm NE}(\omega) \ ,
\end{equation}
a dynamically renormalized distribution 
$\tilde f_0^{\rm NE}= f_0^{\rm NE}(\omega) / \mathcal{N}(\omega)$, with the renormalisation
$\mathcal{N}(\omega)$ given by the sum
of the spectral functions of the leads 
$\Gamma_{L+R}=\sum_{\alpha=L,R}  i(\Sigma_\alpha^>-\Sigma_\alpha^<)$
and of the interaction
$\Gamma_{\rm int}= i(\Sigma^>_{\rm int}-\Sigma^<_{\rm int})$, 
and an extra term $\delta f^{\rm NE}$
corresponding to the inelastic processes given by $\Sigma^<_{\rm int}$, and renormalised by 
the same factor $\mathcal{N}(\omega)$.

The non-interacting distribution $f_0^{\rm NE}$ is formed by two Fermi-Dirac
distributions shifted by the bias $V$. However, the full NE distribution
presents richer features (peaks and dips) characteristics of the electron population
redistribution arising from both the NE and interaction effects.
One can obtain both
accumulation or depletion (i.e. population inversion) in some energy windows, such
features in the NE distribution provide information about the efficiency of 
relaxation/equilibration processes in the system.

Furthermore, another important property of the NE distribution $f^{\rm NE}$ is
related to its functional form.
Indeed, any Feynmann diagrams for the interaction self-energy $\Sigma_{\rm int}$
(taken at any order and for electron-electron e-e or electron-phonon e-ph interaction) 
is expressed in terms of the different electron GFs and phonon GFs. 
The renormalisation of the phonon GFs, if present, is also obtained from another
set of diagrams using the electron GFs (in the case of e-ph interaction).

Since all GFs (either the retarded, the advanced or the lesser, the greater, or 
the (anti)time-order) can be expressed in terms of spectral function $A_C$ alone or in
terms of both the spectral function and the NE distribution, see Eq.(\ref{eq:GF_and_fNE}),
any self-energy is a functional of the spectral functions and of the NE distribution function.
In Appendix \ref{app:fNE_functional}, we show explicitly how such a functional dependence is 
obtained by considering different lowest-order diagrams for the self-energies in case of both
e-e and e-ph interaction.

Therefore, from the general expression Eq.~(\ref{eq:fNE}) defining $f^{\rm NE}$, we can conclude 
that $f^{\rm NE}=f^{\rm NE}[f^{\rm NE}(\omega),A_C(\omega)]$.
The fact that $f^{\rm NE}$ is a functional of itself and of the spectral function permits us to
devise an approach to solve self-consistently the problem by using an iterative scheme. 
Such a scheme is developed in the next section and bears resemblance with conventional 
self-consistent NEGF calculations.

\subsection{Algorithm for NE calculations}
\label{sec:fNE_algo}

The method we present in this section has however some advantages compared to conventional
NEGF calculations.
First of all, we are now dealing with two real functions $f^{\rm NE}(\omega)$ 
and $A_C(\omega)$ instead of complex number functions for the GFs.
More importantly these two functions have well behaved (for numerical purposes) asymptotic limits: 
the spectral function $A_C(\omega)$
has a finite energy-support, i.e. $A_C(\omega) \ne 0$ for $\omega \in [\omega_{\rm min},\omega_{\rm max}]$
otherwise $A(\omega)=0$, and $f^{\rm NE}(\omega)=1$ for $\omega < D^{\rm NE}_\omega$ and 
$f^{\rm NE}(\omega)=0$ for $\omega > D^{\rm NE}_\omega$ where the energy domain $D^{\rm NE}_\omega$ is
roughly the bias window $D^{\rm NE}_\omega = [{\rm min}(\mu_L,\mu_R),{\rm max}(\mu_L,\mu_R)] \pm$ 
several $kT$. 

Hence by using only $f^{\rm NE}(\omega)$ and $A_C(\omega)$, we avoid having to deal with the slow 
decaying behaviour in $1/\omega$ of the real part of the advanced and retarded GFs and self-energies. 
Such slow decay in $1/\omega$ comes from the Fourier transform of the Heavyside function defining the
causality in the retarded (the anti-causality in the advanced)  quantities.
We are not obliged to work with large (i.e. long ranged) energy grids. 
In principle, one should work with a grid larger than $D^{\rm NE}_\omega$ in order to include the 
possible effects of ``hot electrons'' excited well above the bias window due to the interaction.
In practice, we have found that the energy grid could be only the support of the spectral 
function $[\omega_{\rm min},\omega_{\rm max}]$.

\subsubsection{The algorithm}
\label{sec:algo}

The algorithm to perform NE steady state calculations is as follows:

\begin{itemize}
\item[1-]
Start with an initial ($n=0$) spectral function $A^{(n)}(\omega)$ and NE distribution $f^{\rm NE(n)}(\omega)$, 
for example those corresponding to the non-interacting case: $A^{(0)}(\omega)=-\Im m G_0^r(\omega)/\pi$ 
and $f^{{\rm NE}(0)}(\omega)=f_0^{\rm NE}(\omega)$.
\item[2-]
Calculate the corresponding initial self-energies $\Sigma^{\lessgtr(n)}_{\rm int}$ for the chosen model 
of MB interaction. 
\item[3-]
Calculate the next iteration NE distribution $f^{{\rm NE}(n+1)}(\omega)$ from Eq.~(\ref{eq:fNE})
as follows
\begin{equation}
\label{eq:fNE_iter}
f^{{\rm NE}(n+1)} = \frac{f_0^{\rm NE}\Gamma_{L+R} - i \Sigma^{<(n)}_{\rm int} }
{\Gamma_{L+R} + i (\Sigma^{>(n)}_{\rm int} - \Sigma^{<(n)}_{\rm int} ) } \ ,
\end{equation}
with $\Sigma^{\lessgtr(n)}_{\rm int}=\Sigma^\lessgtr_{\rm int}[f^{{\rm NE}(n)},A_C^{(n)}]$.
Note that the quantities $i\Sigma^{\lessgtr(n)}_{\rm int}$ are also real functions.
\item[4-]
Calculate the next iteration spectral function from either

$\bullet$ \emph{method (a)}: using the following expression
\begin{equation}
\label{eq:A_iter}
A_C^{(n+1)}(\omega) = \frac{f^{{\rm NE}(n)}(\omega)}{f^{{\rm NE}(n+1)}(\omega)} A_C^{(n)}(\omega) \ ;
\end{equation}

$\bullet$ \emph{method (b)}: using the definition of the spectral functions 
$i 2 \pi A_C(\omega) = G_C^a(\omega) - G_C^r(\omega) = G_C^< - G_C^> = G_C^r (\Sigma^< - \Sigma^> ) G_C^a$, 
we define
\begin{equation}
\label{eq:A_def_iter}
\begin{split}
& 2\pi A_C^{(n+1)}(\omega)  = \\
& \mathcal{G}^{r(n)} 
\left( \Gamma_{L+R} + i (\hat\Sigma^{>(n)}_{\rm int} - \hat\Sigma^{<(n)}_{\rm int} )
\right)
\mathcal{G}^{a(n)} \ ,
\end{split}
\end{equation}
where $\mathcal{G}^{r/a(n)}$ should be considered as intermediate (dummy) functions defined from
the $n$-th iteration spectral function $A_C^{(n)}(\omega)$ as
\begin{equation}
\label{eq:Gra_iter}
\mathcal{G}^{r/a(n)}(\omega) = \mathcal{H}[\pi  A_C^{(n)}(\omega)] \mp i \pi  A_C^{(n)}(\omega) \ ,
\end{equation}
where $\mathcal{H}[f(x)]$ is the Hilbert transform of function $f(x)$, i.e.
$\mathcal{H}[f(x)]= 1/\pi\ P.V. \int dy\ f(y)/(x-y) $.

And $\hat\Sigma^{\lessgtr(n)}_{\rm int}$ is an intermediate updated version of the self-energy
obtained from 
$\hat\Sigma^{\lessgtr(n)}_{\rm int}=\Sigma^\lessgtr_{\rm int}[f^{{\rm NE}(n+1)},A_C^{(n)}]$.
\item[5-]
Ensure normalisation of $A_C^{(n+1)}$ when using approximated functionals for the NE
distribution such as $f^{\rm NE}_{\rm LOE}$ or $f^{\rm NE}_{(1)}$ (see below).
\item[6-]
Repeat the iteration process, from step 3-, until the desired convergence is achieved 
(either for the NE distribution $f^{{\rm NE}(n+1)}$ or for the spectral function  $A^{(n+1)}(\omega)$
or for both).
\end{itemize}

It should be noted that, similarly to the spectral functions, the spectral ``densities''
of the self-energy of the leads $\Gamma_{L+R}(\omega)$, and of the interaction self-energy
$i (\Sigma^>_{\rm int} - \Sigma^<_{\rm int} )$ are bounded, i.e. there
have zero values outside an energy interval which is roughly the same as $[\omega_{\rm min},\omega_{\rm max}]$.
Hence we do not have to worry about the long-ranged dependence in $1/\omega$ of the real
part of $\mathcal{G}^{r/a(n)}$; and we recover spectral functions $A^{(n+1)}(\omega)$ which
exist only on a finite energy-support.

Furthermore \emph{method (a)} for the calculation of the spectral function presents the great advantage 
of being extremely simple, in comparison to \emph{method (b)} \cite{Note:1}.
However, we have noticed that, in some cases
when the initial spectral function of the iterative loop is too different from the expected result, 
the convergence process might be slower (if not possible at all) for \emph{method (a)} than
for \emph{method (b)}. Hence \emph{method (b)} appears to be more robust upon the choice of the 
initial conditions.
It is entirely possible to combined both schemes in the same algorithm, starting first
with \emph{method (b)} and when some degree of convergence is reached switching to \emph{method (a)} 
to obtain a more accurate level of convergence.

\subsubsection{Approximated NE distributions}
\label{sec:approx_fNE}

The method devised in the previous section can appear at first glance as just another
reformulation of conventional NEGF calculations. One performs calculations 
with another set of two independent (but inter-related) functions: the NE distribution $f^{\rm NE}$ and 
the spectral function $A_C$. 
In conventional NEGF technique, one deals instead with the two independent GFs $G_C^>$ and $G_C^<$. 
As mentioned above, there is indeed a one-to-one correspondence between these two sets of functions.

However, our method offers many advantages and not only on the numerical point of view as explained
previously. 
Indeed, as the NE distribution is a functional of itself and of the spectral function, it 
offers a more direct and natural way of performing approximated calculations, by considering some
specific subsets of inelastic processes.
Such approximations are advantageous to minimize the computational cost of the calculations, which
is an important point for future applications to large and more realistic systems. 

Ultimately $f^{\rm NE}$, given by Eq.~(\ref{eq:fNE}), can be expressed as an infinite series expansion 
in terms of the non-interacting NE distribution $f_0^{\rm NE}$, the spectral
function $A_C$ and the interaction parameters ($\gamma_0$ or $v_q$).
Therefore instead of performing the calculations with the exact expression Eq.~(\ref{eq:fNE}), we can 
always truncate the series expansion to a desired level of accuracy (i.e. selecting a specific 
subset of inelastic processes), suitable for the system and the properties under study.

We provide, in the next section, some example of approximated NE distributions and compare their 
performance against exact calculations using the full $f^{\rm NE}$ distribution. We recall that
the latter are strictly equivalent to NEGF calculations (with the same model of self-energies).

\subsection{Numerical application}
\label{sec:numerics}

We now consider numerical applications for a model of e-ph interacting system,
and we test the different approximations available for the functional forms of the NE
distribution.

\subsubsection{Model of electron-phonon interacting system}
\label{sec:ephmodel}

The Hamiltonian for the region $C$ is
\begin{equation}
\label{eq:H_central}
\begin{split}
  H_C 
  = \varepsilon_0 d^\dagger d + \hbar \omega_0 a^\dagger a +
  \gamma_0 (a^\dagger + a) d^\dagger d,
\end{split}
\end{equation}
where $d^\dagger$ ($d$) is the creation (annihilation) operator of an
electron in the molecular level $\varepsilon_0$. The electron
is coupled, via its charge density $d^\dagger d$, to the vibration mode
(phonon)
of energy $\omega_0$ and the strength of the coupling is given by the constant $\gamma_0$, and
$a^\dagger$ ($a$) creates (annihilates) a quantum in the vibron mode $\omega_0$.

For the transport set-up, the central region $C$ is connected to two ($L$ and $R$) 
one-dimensional semi-infinite tight-binding chains via the hopping integral $t_{0L}$ and
$t_{0R}$. The corresponding $\alpha=L,R$ self-energy is obtained from
the GF at the end of the semi-infinite tight-binding chain and is given by
$\Sigma^r_\alpha(\omega)=t_{0\alpha}^2 e^{{\rm i} k_\alpha(\omega)} /\beta_\alpha $.
A dispersion relation links the energy $\omega$ with the momentum $k_\alpha$ of an
electron in the lead $\alpha$: 
$\omega=\varepsilon_\alpha+2\beta_\alpha \cos k_\alpha(\omega)$. 
The parameters $\varepsilon_\alpha$ and $\beta_\alpha$ are the on-site and off-diagonal
elements of the tight-binding chains.
With such a choice of lead self-energy, we go beyond the wideband limit
(unless $\beta_\alpha$ is much larger than any other parameters). 

The self-energies $\Sigma_{\rm int}$ for the interaction between the electron and 
the vibration mode are calculated using the Born approximation \cite{Dash:2010,Dash:2011}. 
Their expressions are given in Appendix \ref{app:SCBA}.

Finally, in the most general cases, the left and right contacts are different ($\Gamma_L \ne \Gamma_R$) 
and there are asymmetric potential drops, i.e.  $\mu_\alpha=\mu^{\rm eq}+\eta_\alpha V$,
with the condition $\Delta\mu=\mu_L-\mu_R=V$ (i.e. $\eta_L-\eta_R=1$).

\begin{figure}
  \center
  \includegraphics[clip=,width=80mm]{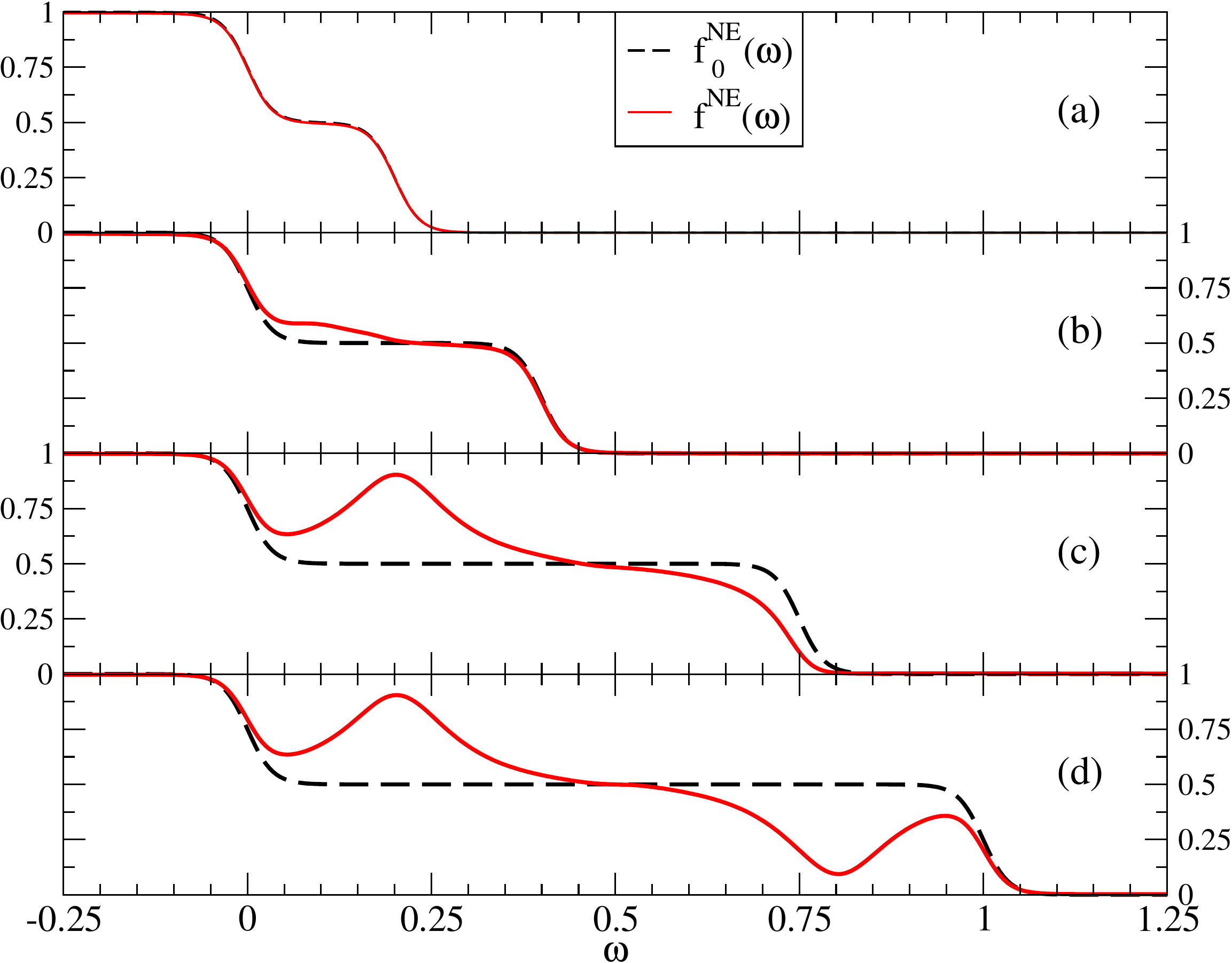}
  \caption{(color online)
NE distribution functions for the off-resonant regime ($\varepsilon_0=0.50$) 
and for different biases $V$. Panel 
(a) $V=0.2<\omega_0$, 
(b) $V=0.4 \sim \omega_0$,
(c) $V=0.75>\omega_0$, 
(d) $V=1.0\gg\omega_0$.
The NE distribution $f^{\rm NE}$ is completely different from the non-interaction
NE distribution $f_0^{\rm NE}$ when $V\ge\omega_0$, in this case inelastic processes 
occur and induce a redistribution
of the electron population in the region $C$.
The other parameters are $\gamma_0=0.09, \omega_0=0.3$,  
$t_{0\alpha}=0.15$, $T_\alpha=0.017$, $\eta_L=1$, 
$\varepsilon_\alpha=0, \beta_\alpha=2$.}
  \label{fig:1}
\end{figure}

\subsubsection{Examples of NE distributions}
\label{sec:examplefNEcalc}

We provide typical examples of the charge redistribution in the
central region induced by both the NE effects and the interaction.
For a given model of interaction self-energies, the full self-consistent calculations
provided by the algorithm in Sec. \ref{sec:algo} are strictly equivalent to 
conventional NEGF calculations.
Hence the results obtained for $f^{\rm NE}$ with our method (and full self-consistency)
are equal to those obtained from NEGF-SCBA calculations \cite{Dash:2010,Dash:2011}.

Figure \ref{fig:1} shows how the NE distribution evolves upon increasing the NE
conditions, i.e. the applied bias, for a typical set of parameters characterising
the off-resonant transport regime.
One can clearly observe the difference between the non-interacting NE distribution
$f_0^{\rm NE}$ and the full distribution $f^{\rm NE}$.
The latter presents features (peaks and dips) which correspond to accumulation or
depletion of the electron population induced by inelastic scattering effects.
Such features are directly related to the peaks in the spectral function.
This single example confirms explicitly that, generally, $f^{\rm NE} \ne f_0^{\rm NE}$ 
as shown analytically in Appendix \ref{app:fNE_withwithout_inter}.

\subsubsection{Approximated NE distributions}
\label{sec:approxfNEcalc}

As mentioned in Section \ref{sec:approx_fNE}, for a given choice of interaction 
self-energies, our approach is fully equivalent to NEGF calculations. 
Both methods corresponds to a partial resummation of a family of diagrams associated 
with the interaction self-energy. 
However, we can further approximate the expression of the NE distribution Eq.~(\ref{eq:fNE}). 
This corresponds to another way of partially resuming the diagrams corresponding 
to $\Sigma_{\rm int}$.

A lowest order expansion (LOE), in terms of the characteristic interaction parameter,
gives an approximated NE distribution  in the following form:
\begin{equation}
\label{eq:fNE_LOE1}
\begin{split}
&f^{\rm NE}(\omega)
\sim 
\left(f_0^{\rm NE}(\omega) - \frac{i \Sigma^<_{\rm int}}{\Gamma_{L+R}}\right)
\left(1 - \frac{i (\Sigma^>_{\rm int}-\Sigma^<_{\rm int} ) }{\Gamma_{L+R}} \right) \\
&\sim
f_0^{\rm NE}\left(1 - i\frac{\Sigma^>_{\rm int}-\Sigma^<_{\rm int} }{\Gamma_{L+R}} \right)
- \frac{i \Sigma^<_{\rm int}}{\Gamma_{L+R}} + \mathcal{O}(\gamma_0^{n\ge 4}) \ .
\end{split}
\end{equation}

Using the expressions for the self-energies $\Sigma^\gtrless_{\rm int}$ given in 
Appendix \ref{app:SCBA}
for the limit $N_{\rm ph}=0$, we find that:
\begin{equation}
\label{eq:fNE_LOE2}
\begin{split}
&f^{\rm NE}_{\rm LOE}(\omega) = f^{\rm NE}_0(\omega) \\ 
&+ \frac{2\pi\gamma_0^2}{\Gamma_{L+R}}\ \left[ 
  A_C(\omega+\omega_0)\ f^{\rm NE}_0(\omega+\omega_0)\ [1-f^{\rm NE}_0(\omega)] \right. \\
&\left. - A_C(\omega-\omega_0)\ [1-f^{\rm NE}_0(\omega-\omega_0)]\ f^{\rm NE}_0(\omega)\ \right] \ ,
\end{split}
\end{equation}
where the terms in $\gamma_0^2$ represent NE inelastic correction terms (to 
the non-interacting distribution $f^{\rm NE}_0$) arising from phonon emission by 
electron and hole.
Such correction terms are proportional to the ratio $\gamma_0^2/\Gamma_{L+R}$, where
$\Gamma_{L+R}$ represents to total escape (injection) rate of electron or hole from (into)
the central region $C$. The order of the interaction parameter is $\gamma_0^2$ as in
lowest order perturbation theory.

Eq.~(\ref{eq:fNE_LOE2}) represents the simplest functional form of the NE distribution
$f^{\rm NE}(\omega) = f^{\rm NE}[f^{\rm NE}_0,A_C](\omega)$.
However, it is a lowest order series expansion in terms of the parameter $\gamma_0$
and is only valid for weak coupling, as we will show below.
It should be noted that the inelastic processes can only occur when the bias $V$ is larger
or equal to the excitation energy, $V\ge\omega_0$, otherwise the factors
associated with phonon emission by 
electron [$f^{\rm NE}_0(\omega+\omega_0) (1-f^{\rm NE}_0(\omega))$] or by 
hole [$f^{\rm NE}_0(\omega) (1-f^{\rm NE}_0(\omega-\omega_0))$] 
are zero over the whole energy range \cite{Note:2}.

Another possible approximation is to consider Eq.~(\ref{eq:fNE}) using only
the non-interacting distribution $f^{\rm NE}_0$ in the evaluation of the 
self-energies $\Sigma^\lessgtr_{\rm int}$.
One then gets
\begin{equation}
\label{eq:fNE_O1}
\begin{split}
f^{\rm NE}_{(1)}(\omega) = &\left[ f^{\rm NE}_0(\omega) \Gamma_{L+R}(\omega) +  \right.\\
& \left. 2\pi\gamma_0^2 A_C(\omega+\omega_0)\ f^{\rm NE}_0(\omega+\omega_0)
\right] / \mathcal{N}(\omega) \ ,
\end{split}
\end{equation}
with $\mathcal{N}(\omega)= \Gamma_{L+R}(\omega) + \Gamma_{\rm int}(\omega)$
and 
\begin{equation}
\label{eq:GammaInt_O1}
\begin{split}
\Gamma_{\rm int}(\omega) = 2\pi\gamma_0^2 
&\left[   A_C(\omega-\omega_0)\ [ 1- f^{\rm NE}_0(\omega-\omega_0)] \right. \\
&\left. + A_C(\omega+\omega_0)\ f^{\rm NE}_0(\omega+\omega_0) \right] . 
\end{split}
\end{equation}

\begin{figure}
  \center
  \includegraphics[clip=,width=72mm]{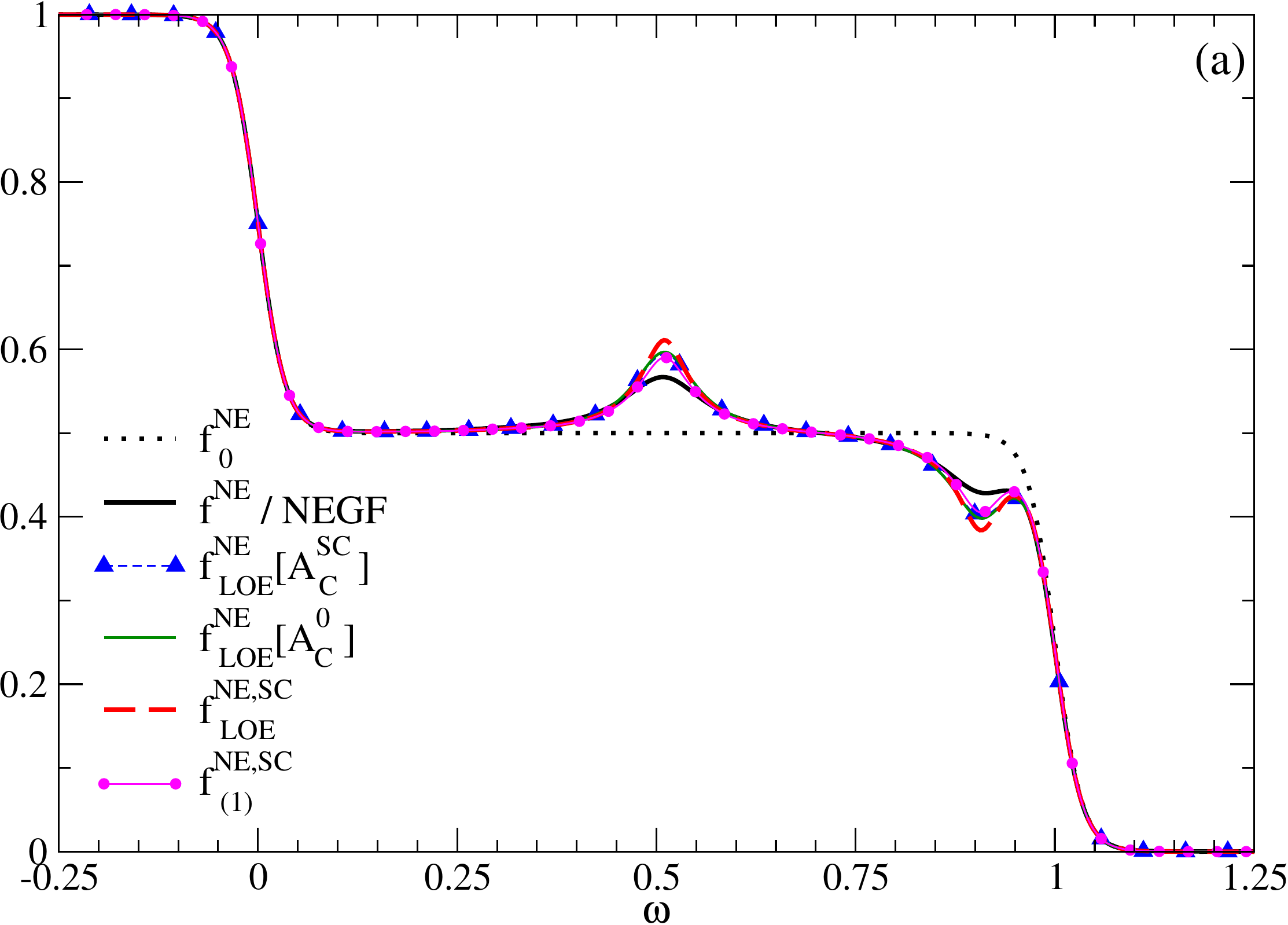}
  \includegraphics[clip=,width=72mm]{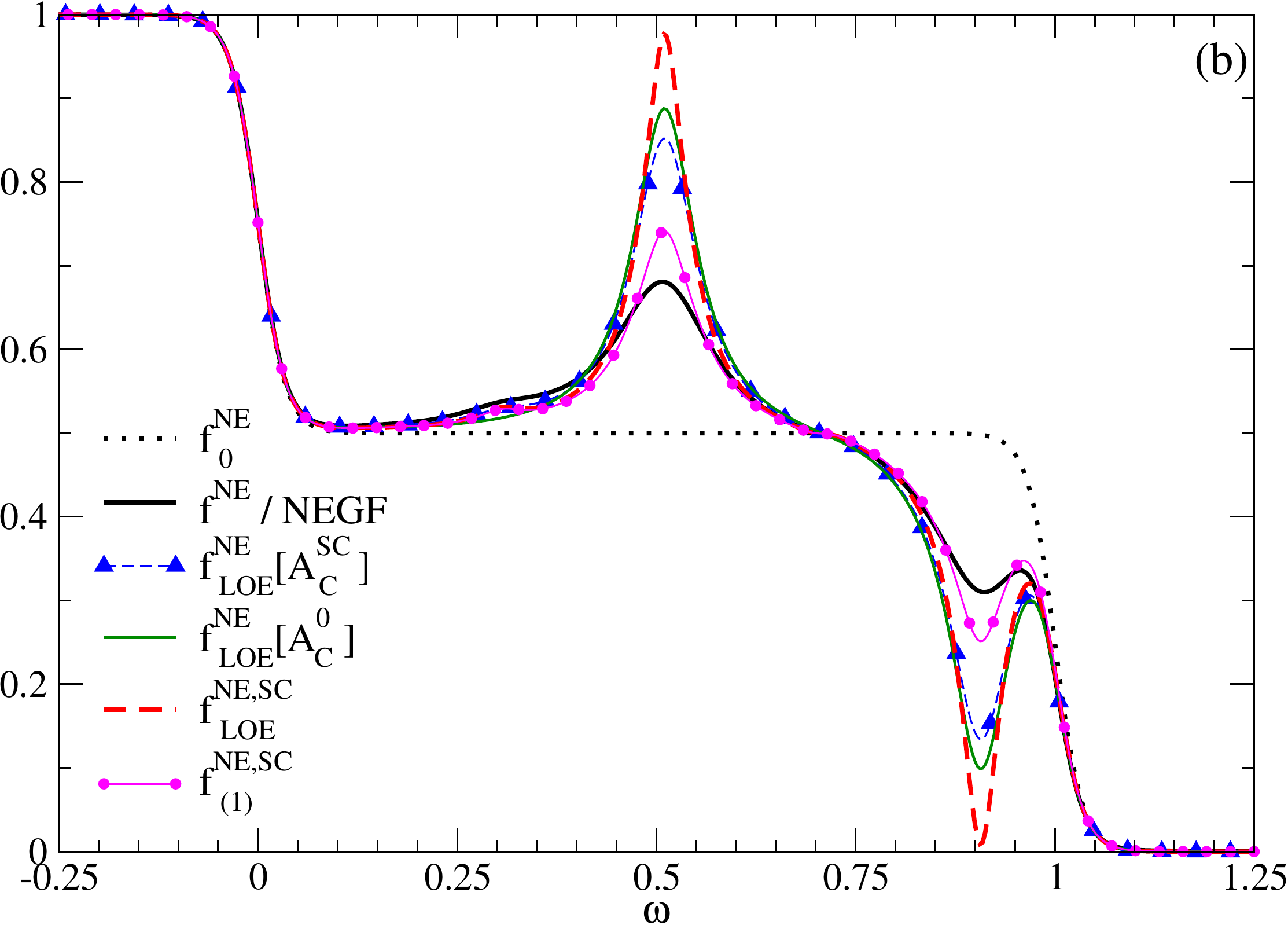}
  \includegraphics[clip=,width=72mm]{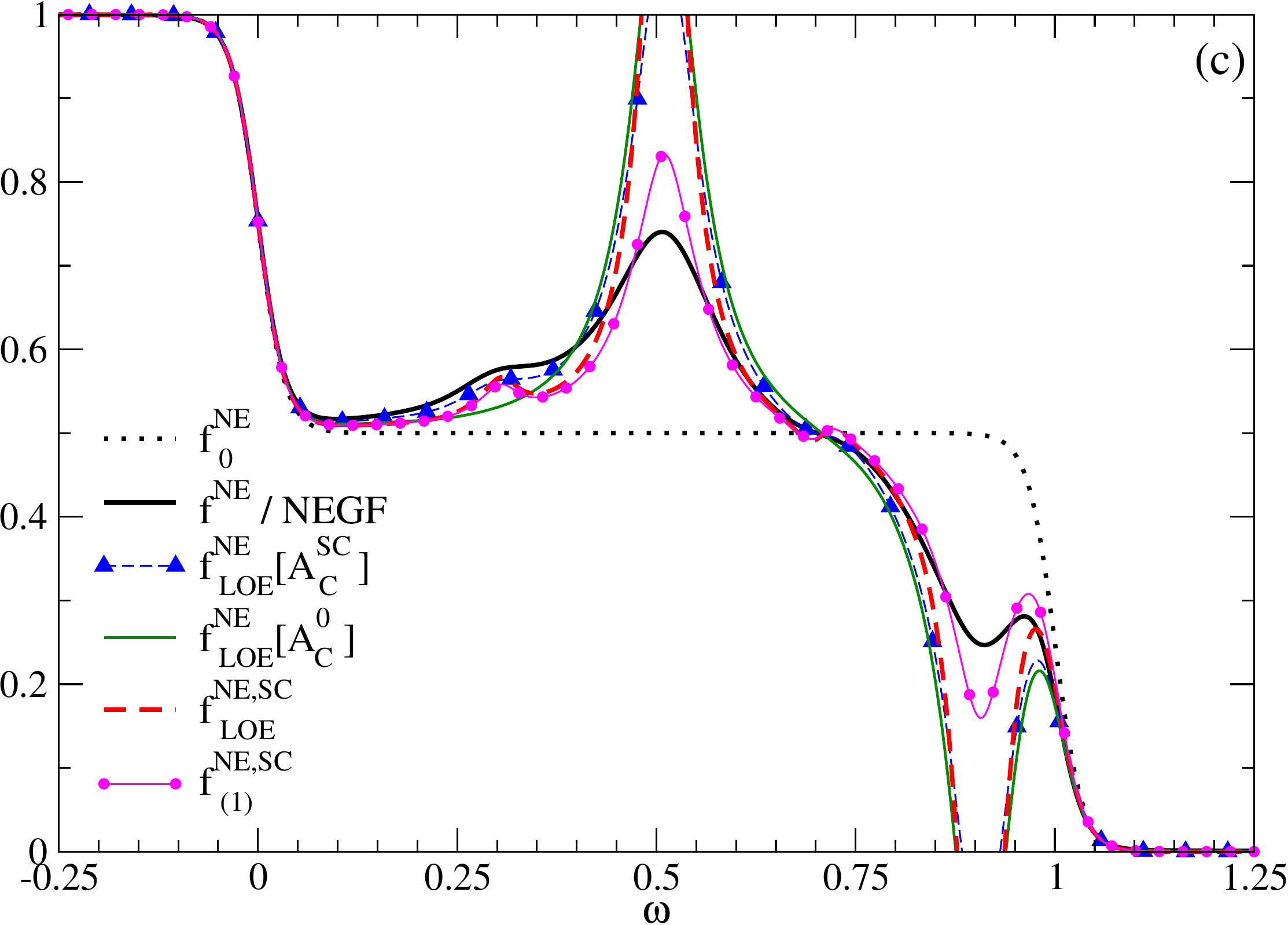}
  \caption{(color online)
NE distribution functions for the off-resonant regime ($\varepsilon_0=0.70$) 
for different approximations and for different coupling strengths $\gamma_0$. 
Panel 
(a) $\gamma_0=0.03$ ($\gamma_0/\omega_0=0.15$), 
(b) $\gamma_0=0.06$ ($\gamma_0/\omega_0=0.3$), 
(c) $\gamma_0=0.08$ ($\gamma_0/\omega_0=0.4$). 
Only for weak coupling, all approximated NE distributions provides a good
representation of the exact distribution $f^{\rm NE}$. 
We recall that for fully self-consistent calculations, the results obtained
for $f^{\rm NE}$ with our method are strictly equivalent to those obtained
from NEGF calculations.
See text for more
detailed comments.
The other parameters are $V=1.0$, $\omega_0=0.2$,  
$t_{0\alpha}=0.22$, $T_\alpha=0.017$, $\eta_L=1$, 
$\varepsilon_\alpha=0, \beta_\alpha=2$.}
  \label{fig:2}
\end{figure}

In figure \ref{fig:2}, we show different NE distributions calculated with different
approximations: the non-interaction distribution $f^{\rm NE}_0$, the full
self-consistent distribution $f^{\rm NE}$, different approximations for
the LOE distribution $f^{\rm NE}_{\rm LOE}$ and $f^{\rm NE}_{(1)}$. 
$f^{\rm NE}_{\rm LOE}[A_C^{\rm SC}]$ is calculated from Eq.~(\ref{eq:fNE_LOE2})
using the full self-consistent spectral function $A_C^{\rm SC}(\omega)$,
$f^{\rm NE}_{\rm LOE}[A_C^0]$ is calculated from Eq.~(\ref{eq:fNE_LOE2})
using the non-interacting spectral function $A_C^0(\omega)$.
Finally $f^{\rm NE,SC}_x$ is obtained from a self-consistent calculation for the 
spectral function (see Sec.~\ref{sec:algo}) using the functional form Eq.~(\ref{eq:fNE_LOE2})
for $x=$LOE, or Eq.~(\ref{eq:fNE_O1}) for $x=(1)$.

One can see in the upper panel (a) of Figure \ref{fig:2} that, for weak e-ph coupling, 
any approximations for  $f^{\rm NE}_{\rm LOE}$ give the same results,
and provide a good representation of  the exact $f^{\rm NE}$.
The distribution $f^{\rm NE,SC}_{(1)}$ provides a better representation for $f^{\rm NE}$.
The amplitude of $f^{\rm NE}_{\rm LOE}$ is slightly different from $f^{\rm NE}$, because
it is obtained from a series expansion and is not fully renormalised
by the factor $\mathcal{N}(\omega)$
Such a renormalisation is however included in $f^{\rm NE,SC}_{(1)}$.

For larger e-ph coupling, the difference between $f^{\rm NE}_{\rm LOE}$ 
and $f^{\rm NE}$ increases, as can be expected from any perturbation series expansion.
The LOE gives physical results only when the electron-phonon coupling is such as
$2\pi\gamma_0^2/\Gamma_{L+R} \ {\rm max}[A_C(\omega)] < 0.5$. Otherwise one gets
non-physical results for the NE distribution, i.e. $f^{\rm NE}_{\rm LOE} > 1$ or 
$f^{\rm NE}_{\rm LOE} < 0$,
as shown in the lower panel (c) of Figure \ref{fig:2}.
Such a behaviour never occurs for the distribution $f^{\rm NE,SC}_{(1)}$ since it
contains the proper renormalisation.
Therefore, in general, it is better to use an approximated distribution like $f^{\rm NE}_{(1)}$
than the LOE.

Another important point to mention is shown in the panels (b) and (c) in Figure \ref{fig:2}:
the inelastic processes (see side band peak located around $\omega\sim 0.3$) are
only reproduced in the LOE when some form of self-consistent has been used,
i.e. either in the form of $f^{\rm NE}_{\rm LOE}[A_C^{\rm SC}]$ or $f^{\rm NE,SC}_{\rm LOE}$.
The LOE distribution calculated with the non-interacting spectral function $A_C^0(\omega)$
is not able to reproduce such effects.

Finally it should also be noted that all the self-consistent calculations including 
approximated distributions, like  $f^{\rm NE}_{\rm LOE}$ or $f^{\rm NE}_{(1)}$,
converge much more faster than the full calculation for $f^{\rm NE}$ 
(see footnote [\onlinecite{Note:3}] for more details). 
Such a numerical improvement is important for the calculations of more realistic 
and larger systems.

\subsubsection{Current and IETS signal}
\label{sec:IVandIEST}

\begin{figure}
  \center
  \includegraphics[clip=,width=75mm]{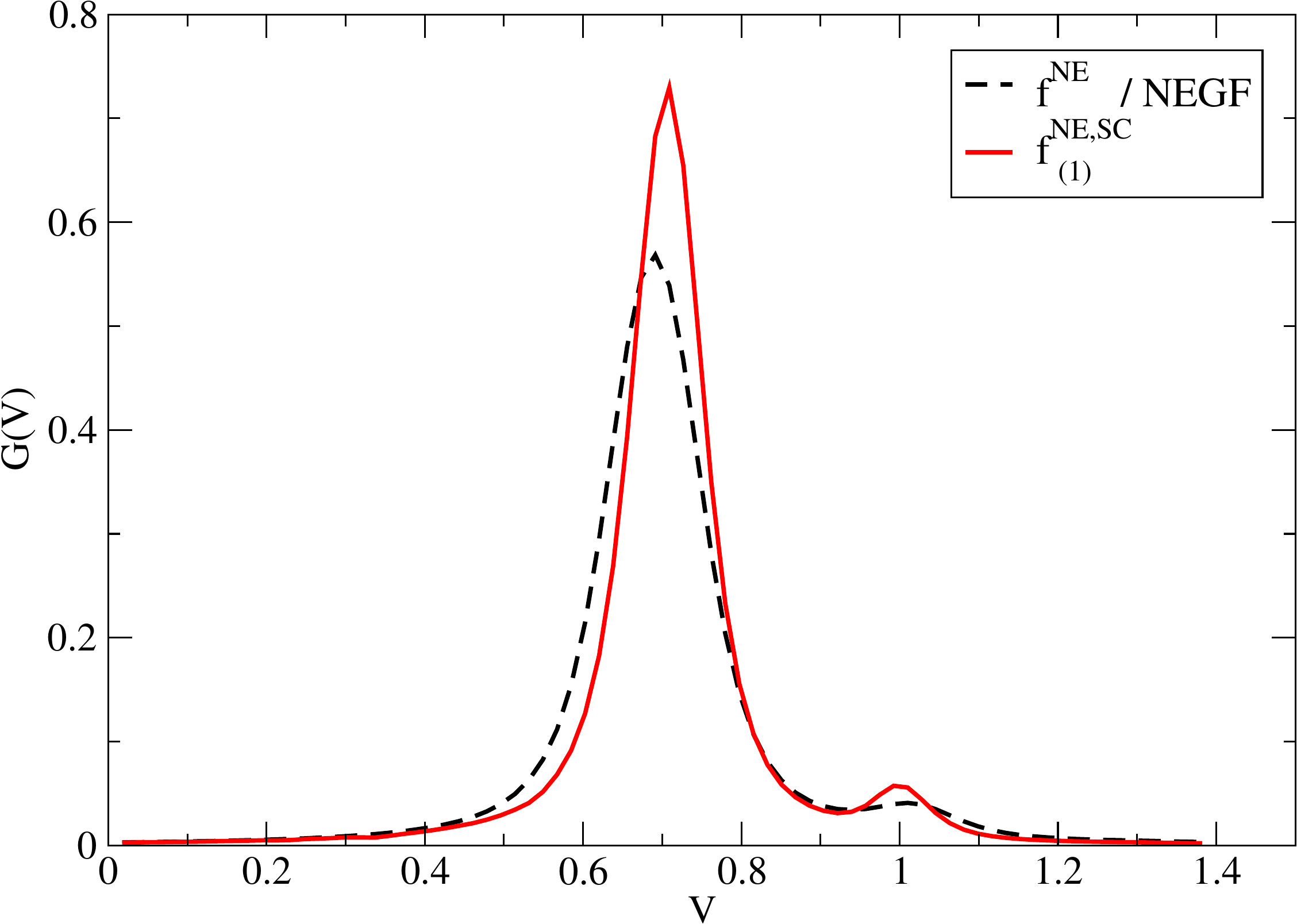}
  \caption{(color online)
Dynamical onductance $G(V)=dI/dV$ (in unit of quantum of conductance $G_0$) 
for the off-resonant regime ($\varepsilon_0=0.70$).
$G(V)$ calculated with the approximated distribution $f^{\rm NE}_{(1)}$ (full line) 
gives a good representation of the conductance calculated with the exact 
distribution $f^{\rm NE}$ (broken line). The latter is strictly equivalent
to NEGF calculations.
The other parameters are $\omega_0=0.3$, $\gamma_0=  0.10$,
$t_{0\alpha}=0.19$, $T_\alpha=0.017$, $\eta_L=1$, 
$\varepsilon_\alpha=0, \beta_\alpha=2$.}
  \label{fig:3}
\end{figure}

Figure \ref{fig:3} shows a typical result for the dynamical conductance $G(V)=dI/dV$
obtained in the off-resonance transport regime.
The current is calculated as in Ref.~[\onlinecite{Ness:2010}] using different approximations 
for the NE distribution function.

The conductance $G(V)$ calculated with the approximated distribution $f^{\rm NE}_{(1)}$ 
provides a good representation of the conductance calculated with the exact 
distribution $f^{\rm NE}$. The peak position are well reproduced, but the amplitude
of the conductance peaks is slightly larger with $f^{\rm NE}_{(1)}$.
This is due to the lack of full renormalisation of $f^{\rm NE}_{(1)}$ in
comparison to $f^{\rm NE}$. The approximated distribution $f^{\rm NE}_{(1)}$ always gives 
a slightly larger electron population as shown in Fig.~\ref{fig:2}.

We do not show the results obtained with $f^{\rm NE}_{\rm LOE}$ since for
coupling strengths $\gamma_0/\omega_0 > 0.3$, $f^{\rm NE}_{\rm LOE}$ gives
non-physical results as shown in panel (c) of Figure \ref{fig:2}.

\begin{figure}
  \center
  \includegraphics[clip=,width=75mm]{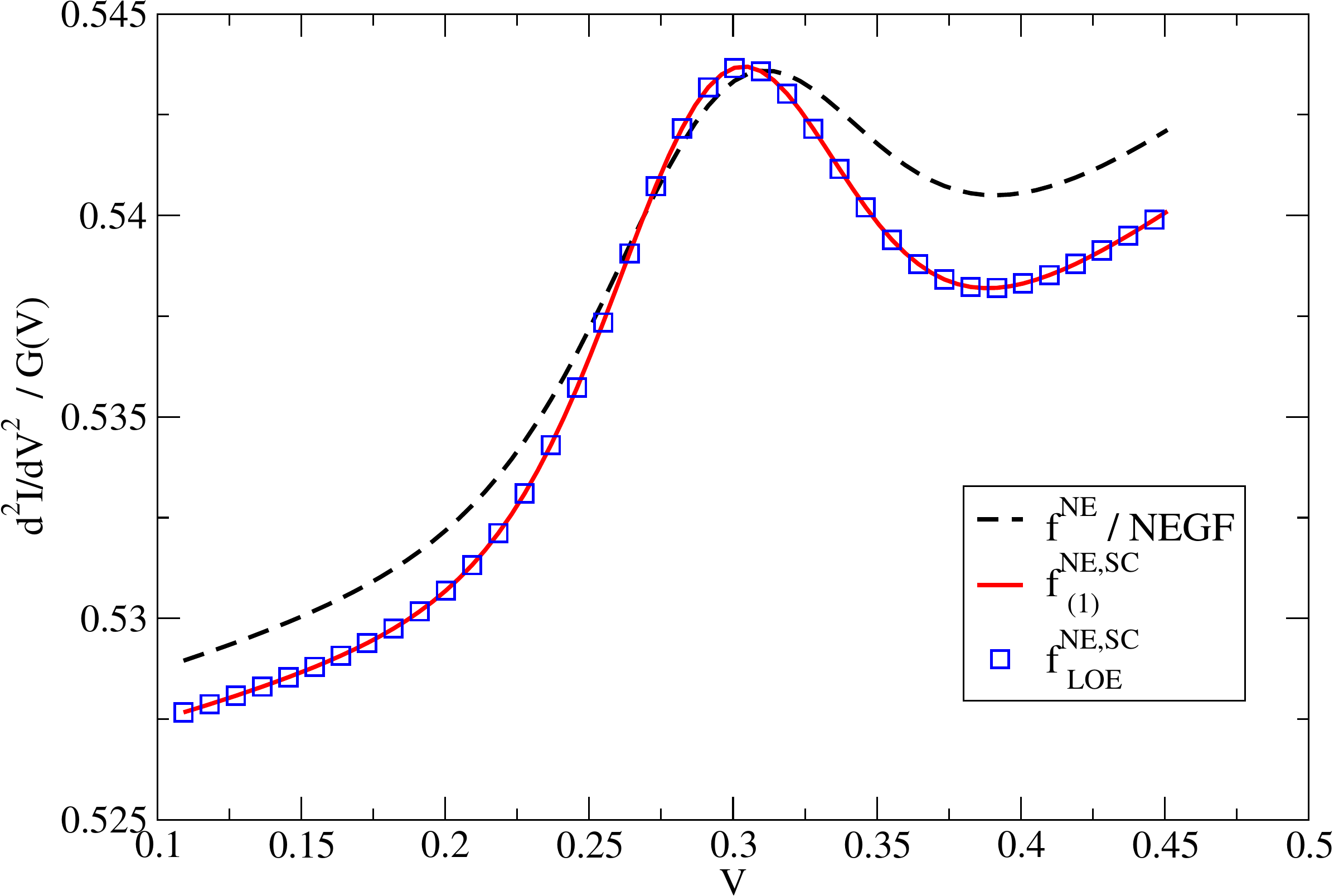}
  \caption{(color online)
IETS signal $d^2I/dV^2$, normalised by the conductance $G(V)$, 
for the far-off-resonant regime ($\varepsilon_0=3.70$).
The IETS calculated with the approximated distributions $f^{\rm NE}_{(1)}$ (full line)
or even  $f^{\rm NE}_{\rm LOE}$ (squares)
gives a good representation of the IETS calculated with the exact 
distribution $f^{\rm NE}$ (broken line). The results obtained with $f^{\rm NE}$
are strictly equivalent to NEGF calculations.
The other parameters are $\omega_0=0.3$, $\gamma_0=0.12$,
$t_{0L}=0.45$, $t_{0R}=0.10$, $T_\alpha=0.017$, $\eta_L=1$, 
$\varepsilon_\alpha=0, \beta_\alpha=2$.}
  \label{fig:4}
\end{figure}

The inelastic effects are best represented by the inelastic electron tunnelling
spectra (IETS) provided from the second derivative of the current versus the
applied bias.
Figure \ref{fig:4} shows such a signal normalised by the conductance.
As expected for the off-resonance regime \cite{Galperin:2004b,Frederiksen:2004,Frederiksen:2007,Dash:2011}, 
we obtain a peak in the IETS for the voltage threshold $V\sim\omega_0$.
The exact IETS signal calculated with the distribution $f^{\rm NE}$ is well
presented by the IETS calculated with the approximated distribution $f^{\rm NE}_{(1)}$.

More interestingly, the results obtained with the LOE approximated distribution 
$f^{\rm NE}_{\rm LOE}$ also give a good representation of the IETS signal, even
for a coupling strength $\gamma_0/\omega_0=0.4$.
We interpret such a behaviour in the following manner: for small applied bias, where
the transport is mostly tunneling and away from any resonant transport 
mechanisms, the LOE distribution $f^{\rm NE}_{\rm LOE}$ is still realistic 
(i.e. $0<f^{\rm NE}_{\rm LOE}<1$) and quite close to the distribution $f^{\rm NE}_{(1)}$.
Hence both distributions provide similar results for the IETS signal.

However, whenever the bias is large enough to include any resonances (main resonance
or any phonon-side band peak), the LOE distribution will provide a non-physical
behaviour as shown in panel (c) of Figure \ref{fig:2}.

\section{Towards more complex systems}
\label{sec:realistic}

In order to extend the previous results to more realistic systems, we need to include 
several electron states and eventually several vibration modes in
the central region.
For that , we follow the same reasoning as in section ~\ref{sec:onelevel}, and
consider the GFs as being matrices $G_{nm}(\omega)$ in the electron level (or site) 
representation. The self-energies are also matrix in such a representation. 
We then define a new matrix for the NE distribution $f^{\rm NE}_{nm}$ as follows:
\begin{equation}
\label{eq:GFfNEnm}
G^<_{C,nm}(\omega)  = - \sum_l f^{\rm NE}_{nl}(\omega)    \left[ G^r_{C,lm}(\omega) - G^a_{C,lm}(\omega) \right] \ .
\end{equation}
With a few lines of algebra, we find that
\begin{equation}
\label{eq:fNEnm}
\begin{split}
&\underline{f}^{\rm NE}(\omega) = \underline{G}^<_C \left[ \underline{G}^a_C - \underline{G}^r_C \right]^{-1} \\
&= \underline{G}^r_C \underline{\Sigma}^< \underline{G}^a_C
\left[ \underline{G}^r_C\ [(\underline{G}^r_C)^{-1} - (\underline{G}^a_C)^{-1}]\ \underline{G}^a_C \right]^{-1} \\
&= \underline{G}^r_C \underline{\Sigma}^< [\underline{\Sigma}^a - \underline{\Sigma}^r]^{-1} (\underline{G}^r_C )^{-1} \\
&= \underline{G}^r_C 
\left( \underline{f}_0^{\rm NE} \underline{\Gamma}_{L+R} - i \underline{\Sigma}_{\rm int}^<
\right)
\left[ \underline{G}^r_C \left( \underline{\Gamma}_{L+R} + \underline{\Gamma}_{\rm int} \right) \right]^{-1} \ ,
\end{split}
\end{equation}
where $\underline{\Gamma}_{\rm int} = i \left( \underline{\Sigma}_{\rm int}^> - \underline{\Sigma}_{\rm int}^< \right)$.

The equation Eq.~(\ref{eq:fNEnm}) for $f^{\rm NE}_{nm}$ is more complicated than Eq.~(\ref{eq:fNE}) 
because of the presence of the retarded GF terms which do not cancel in the general matrix form.
Furthermore, the physical interpretation of $\underline{f}^{\rm NE}$ is more complicated. However
the diagonal matrix elements $f^{\rm NE}_{nn}$ represent the occupations of the level $n$, and the 
off-diagonal matrix elements represent some form of probability rate of transition between states. 

It should noticed that, however, all the functional analysis we have performed in Section 
\ref{sec:fNE} and Appendix \ref{app:fNE_functional} still hold for the matrix case, 
i.e. the interaction self-energy is functional of the spectral function and of the NE distribution. 
Furthermore, $\underline{G}^r_C$ is also a functional of the spectral function, $A_{C,nm}$, which 
is now given in a matrix form, and
$G^r_{C,nm}(\omega) = \mathcal{H}[\pi  A_{C,nm}(\omega)] - i \pi  A_{C,nm}(\omega)$.

Hence we can still use the functional property of the NE distribution, that is  
$\underline{f}^{\rm NE}=\underline{f}^{\rm NE}[\underline{f}^{\rm NE}(\omega),\underline{A}_C(\omega)]$,
to devise a self-consistently iterative scheme to solve the problem. 
However now, we have to take into account all the different matrix elements of the NE distribution 
and spectral functions. 

We can choose for convenience that the coupling of the central region $C$ to the lead $\alpha$ is diagonal
in the $n,m$ representation: $\Gamma_{\alpha nm} = \Gamma_{\alpha,n} \delta_{nm}$. Hence the non-interacting
NE distribution $\underline{f}_0^{\rm NE}$ is diagonal as well, with matrix elements:
\begin{equation}
\label{eq:f0NEn}
\begin{split}
{f}_{0,n}^{\rm NE} = 
\frac{ f_{L,n} \Gamma_{L,n}(\omega) + f_{R,n} \Gamma_{R,n}(\omega) }
     { \Gamma_{L,n}(\omega) + \Gamma_{R,n}(\omega) } \ ,
\end{split}
\end{equation}
with $f_{\alpha,n}=f_\alpha(\omega=\epsilon_n)$ the population of the eigenvalue $\epsilon_n$ of state $n$
given from the statistics of the lead $\alpha$.

Furthermore if the interaction is such that $\underline{\Sigma}_{\rm int}^\lessgtr$ is also diagonal, 
the terms in $G^r$ cancel in Eq.~(\ref{eq:fNEnm}); and we end
up with a set of $n=1,..N$ equations like Eq.~(\ref{eq:fNE}) for $f_{nn}^{\rm NE}(\omega)$ which need to be
solve self-consistently for the $n$ distributions and the $n$ spectral functions $A_{C,nn}(\omega)$.

However, in the most general cases, $\underline{\Sigma}_{\rm int}^\lessgtr$ is not diagonal, and one would
need to solve the problem in a matrix form.
For example, a generalisation of the self-energies for electron-phonon coupling, given in Appendix \ref{app:SCBA},
would be \cite{Frederiksen:2004,Viljas:2005,Yamamoto:2005,Frederiksen:2007,Asai:2008,Arroyo:2010,Rossen:2013} 
\begin{equation}
\label{eq:SEnm}
\Sigma_{{\rm int},nm}^\lessgtr(\omega) = \sum_\nu i \int \frac{d u}{2\pi} 
    D_{0,\nu}^\lessgtr(u) \gamma_{\nu,np} G^\lessgtr_{C,pq}(\omega - u) \gamma_{\nu,qm}
\end{equation}
where the coupling matrix elements $\gamma_{\nu,np}$ correspond to an excitation of the vibration
mode $\nu$ (emission or absorption of a quantum) with electronic transition between state $n$ and $p$.

We provide in Appendix \ref{app:2levelmodel} a specific example of a two-level model coupled to
different vibration modes and show how to calculate the different matrix elements 
of $\underline{f}^{\rm NE}$.

As far as we know, calculations for realistic systems (i.e. several electron levels and 
vibration modes) have not yet been performed for the full range of NE and MB effects.
NE distribution functions have been used in large systems but only for non-interacting
cases or for cases where the interactions are treated in a mean-field manner \cite{Louis+Palacios:2003}.  
The effects of NE and MB effects for e-ph coupled realistic systems have been
considered, however only at the level of a lowest order expansion for the coupling, and
in conjunction with some form of self-consistency \cite{Frederiksen:2007,Asai:2008,Arroyo:2010,Rossen:2013}.

The really important point in the use of NE distributions for complex systems is that
both the NE and MB effects are taken into account in the statistics of the finite size
open quantum system (the central region $C$). 
The NE distributions give the (fractional) electron population
in the corresponding electronic levels in the presence of the NE conditions and for
a given model of the MB effects (self-energies).
One could envisage incorporating such NE statistics in density-functional-based
codes able to deal with fractional occupation numbers for the corresponding Kohn-Sham
states.

\section{Conclusion}
\label{sec:ccl}

We have developed an alternative scheme to calculate the non-equilibrium (steady state) properties 
of open quantum systems. The method is based on the use of NE distribution and spectral functions.
The method presents several advantages, but is strictly equivalent to conventional steady-state
NEGF calculations, when using the same level of approximations for the MB interaction.
This is because there is a one-to-one correspondence with the NE distribution 
and spectral functions and the different GFs used in NEGF.
The advantages of our method resides in the fact that the NE distribution and spectral 
functions have well behaved features for numerical applications,
and that, for the single level model, one works with purely real-number quantities. 

Furthermore, our approach offers the possibility to introduce further approximations, 
not only at the level of the MB interaction (as in NEGF), but also at the level of 
the functional forms used for the NE distributions.
Introducing approximations at this level is important to reduce the computational
cost of the method. For the model of electron-phonon coupled system we have
studied, such approximations provide a good representation of the full exact
results, for either the NE distributions themselves or for physical measurable 
quantities such as the conductance and the IETS signal. 
An extension to systems consisting of several electron levels and several vibration 
modes has also been provided.

The concept of NE distribution functions also give more direct 
physical information about the open quantum system, for example in terms of
depletion or accumulation of the electron population induced by the NE and MB 
effects. The NE distribution is also a useful concept to study other properties
of the open quantum system such as the NE charge susceptibility \cite{Ness:2012} and
the NE fluctuation-dissipation relations \cite{Ness_NEFD:2013}.

We expect that such a method will be useful for the study of large and more
realistic systems \cite{Dash:2012}, such a single-molecule thermoelectric devices,
as some approximated version of the NE distributions could be implemented in 
density-functional-based calculations \cite{Note:4}.

\begin{acknowledgments}
HN warmly thanks L.K. Dash for her precious comments and for her NEGF-SCBA
code which served as the basis and as a reference tool for the present study.
HN acknowledges engaging discussions with T.N. Todorov on distribution functions 
for quantum transport.
\end{acknowledgments}

\appendix

\section{Interacting versus non-interacting NE distributions}
\label{app:fNE_withwithout_inter}

From the general expression of $f^{\rm NE}(\omega)$ in Eq.~(\ref{eq:fNE}), there is no \emph{a priori} 
reason for $f^{\rm NE}$ to be equal to the NE non-interacting distribution $f_0^{\rm NE}$.

In the very special cases where the interaction self-energy $\Sigma^<_{\rm int}$ follows the
non-interacting statistics, i.e. in the sense that
\begin{equation}
\Sigma^<_{\rm int} \stackrel{?}{=}
  -f_0^{\rm NE}(\Sigma^r_{\rm int}-\Sigma^a_{\rm int}) = 
  -f_0^{\rm NE}(\Sigma^>_{\rm int}-\Sigma^<_{\rm int}) \ ,
\end{equation}
we obtain straightforwardly from Eq.~(\ref{eq:fNE}) that $f^{\rm NE}=f_0^{\rm NE}$. 
Then all quantities, GFs and self-energies, follow the statistics given by the non-interacting case, 
as suggested in Ref.~[\onlinecite{Kirchner:2013}].
However this is generally not true.

Indeed, even when the interactions are present only in the central region, it is not
possible to ignore their indirect MB effects which spread throughout the systems.
Such effects need to be incorporated into the local statistics. 
The latter cannot simply arise from the (non-interacting) leads statistics only.

For example, in the Anderson impurity model, the Kondo cloud generated by electron-electron interaction 
expands over more than the single site where the interaction is present. 
For electron-phonon interaction, when one performs a Lang-Firsov unitary transformation to diagonalise
the interacting part of the Hamiltonian, one needs to keep the effects of such a transformation
onto the effective coupling matrix elements between the (now diagonal) central region
and the leads' Hamiltonians. In simple words, one could say that the electron-phonon interaction
is now crossing at the contacts between the central region and the leads. Therefore, there is no
reason to assume that the corresponding statistics would be given by the non-interacting one.

Moreover, there are clear indications from numerical calculations that $f^{\rm NE} \ne f_0^{\rm NE}$.
This has been shown for electron-electron interaction 
(for example, see Figure 3 in Ref.~[\onlinecite{Hershfield:1991}])
and for electron-phonon interaction
(for example, see Figure 5 and 6 in Ref.~[\onlinecite{Lake:1992}],
Figure 6 in Ref.~[\onlinecite{Kral:1997}] and Figure 7 in Ref.~[\onlinecite{Koch:2011}]). 
We also provide a few examples in Section \ref{sec:examplefNEcalc}.

We can also convince ourself that generally $f^{\rm NE} \ne f_0^{\rm NE}$ by considering the following
example for electron-phonon interaction.
The lowest order diagram for which the self-energies $\Sigma^\lessgtr_{\rm int}$ are not vanishing is 
the Fock diagram \cite{Dash:2010,Dash:2011} (see Appendix \ref{app:SCBA}):
\begin{equation}
\label{eq:SigmaFock}
\begin{split}
\Sigma_{\rm int}^{F,\lessgtr}(\omega) =  \gamma_0^2  \left[
N_{\rm ph} G_C^\lessgtr(\omega \mp \omega_0)
+ (N_{\rm ph} + 1) G_C^\lessgtr(\omega \pm \omega_0) \right] . 
\end{split}
\end{equation}
One can use the ratio  $\Sigma_{\rm int}^{F,>}/\Sigma_{\rm int}^{F,<}$ to define 
a distribution function 
\begin{equation}
\label{eq:fNEint}
f^{\rm NE}_{\rm int}(\omega)
=\left( 1 - \frac{\Sigma_{\rm int}^>(\omega)}{\Sigma_{\rm int}^<(\omega)} \right)^{-1}
\end{equation}
such as $\Sigma^<_{\rm int}=-f^{\rm NE}_{\rm int}(\Sigma^>_{\rm int}-\Sigma^<_{\rm int})=
-f^{\rm NE}_{\rm int}(\Sigma^r_{\rm int}-\Sigma^a_{\rm int})$.
At low temperature $N_{\rm ph}=0$ and the ratio  
\begin{equation}
\label{eq:SigmaFock_ratio1}
\frac{\Sigma_{\rm int}^{F,>}}{\Sigma_{\rm int}^{F,<}}=\frac{G_C^>(\omega - \omega_0)}{G_C^<(\omega + \omega_0)}
\end{equation}
defines a distribution $f^{\rm NE}_{\rm int}$ which is clearly different 
from $f_0^{\rm NE}$. 
Indeed if $f^{\rm NE}_{\rm int}=f_0^{\rm NE}$, one has
$\Sigma_{\rm int}^{F,>}/\Sigma_{\rm int}^{F,<}= (f_0^{\rm NE}-1)/f_0^{\rm NE}$ which is not possible
from the definition of Eq.~(\ref{eq:SigmaFock_ratio1}).

To further convince ourselves, let assume that $G_C^\lessgtr$ were following the distribution $f_0^{\rm NE}$.
Then from Eq.~(\ref{eq:SigmaFock_ratio1}), we would have
\begin{equation}
\label{eq:SigmaFock_ratio2}
\begin{split}
\frac{\Sigma_{\rm int}^{F,>}}{\Sigma_{\rm int}^{F,<}}
& = 
\frac{G_C^>(\omega - \omega_0)}{G_C^<(\omega + \omega_0)}
=
\frac{f_0^{\rm NE}(\omega - \omega_0)-1}{f_0^{\rm NE}(\omega + \omega_0)} \
\frac{A_C(\omega - \omega_0)}{A_C(\omega + \omega_0)} \\
& \ne
\frac{f_0^{\rm NE}(\omega)-1}{f_0^{\rm NE}(\omega)} \ ,
\end{split}
\end{equation} 
where $A_C(\omega)$ is the spectral function of the central region $C$.
The inequality in Eq.~(\ref{eq:SigmaFock_ratio2}) holds even for the symmetric electron-hole case
\cite{Note:5}.

Hence, we can safely conclude that, in the most general cases, the two distribution functions 
$f^{\rm NE}$ and $f_0^{\rm NE}$ differ from each other.

\section{The electron-phonon self-energies}
\label{app:SCBA}

The electron-phonon self-energies in the central region $C$ are calculated within the self-consistent
Born approximation. 
The details of the calculations are reported elsewhere \cite{Dash:2010,Dash:2011} so
we briefly recall the different expressions for the self-energies
$\Sigma^{x}_{\rm int}(\omega)=\Sigma^{H,x}_C(\omega) + \Sigma^{F,x}_C(\omega)$ with
\begin{equation}
\label{eq:SigmaC_Hartree}
\begin{split}
  \Sigma^{H,r}_C =  \Sigma^{H,a}_C =  2 \frac{\gamma_0^2}{\omega_0} 
\int \frac{d\omega^\prime}{2\pi} i G^<_C(\omega^\prime) 
 =  - 2 \frac{\gamma_0^2}{\omega_0} \langle n_C \rangle \ ,
\end{split}
\end{equation}
with $\langle n_C \rangle = - i \int {d\omega}/{2\pi}\  G^<_C(\omega)$ 
and
\begin{equation}
\label{eq:SigmaClessgrt_Fock}
\Sigma^{F,\lessgtr}_C(\omega)  =  i \gamma_0^2 
\int \frac{d u}{2\pi} \
    D_0^\lessgtr(u) \
    G^\lessgtr_C(\omega - u) \ ,
\end{equation}
and
\begin{equation}
\label{eq:SigmaCr_Fock}
\begin{split}
\Sigma^{F,r}_C(\omega)  = i \gamma_0^2 
      \int \frac{d u}{2\pi} & D_0^r(\omega - u)
      \left( G^<_C(u) + G^r_C(u) \right)  \\ 
+ & D_0^<(\omega -u) G^r_C(u) \ ,
\end{split}
\end{equation}
with the usual definitions for the bare vibron GF $D_0^x$:
\begin{equation}
\label{eq:D0}
  \begin{split}
    D_0^\lessgtr(\omega) & = -2\pi i \left[ N_{\rm ph} 
\delta(\omega \mp \omega_0) + (N_{\rm ph} + 1) \delta(\omega \pm \omega_0) \right] \\
    D_0^r(\omega) & = \frac{1}{\omega - \omega_0 +i 0^+} 
    - \frac{1}{\omega + \omega_0 +i 0^+} \ , 
  \end{split}
\end{equation}
where $N_{\rm ph}$ is the averaged number of excitations in
the vibration mode of frequency $\omega_0$ given by the Bose-Einstein 
distribution at temperature $T_{\rm ph}$. 

We are mostly interested to see how the inelastic effects are reproduced
by our method based on the NE distribution. Therefore we ignore the contribution of
the static part of the interaction, i.e. the Hartree-like self-energy 
$\Sigma^{H,r/a}_C$, in the calculations. Note however that since the NE distribution
is defined from the lesser and greater components of the interaction self-energies,
the Hartree-like component is not relevant for the calculation of $f^{\rm NE}$.

\section{Functional forms of the NE distribution}
\label{app:fNE_functional}

We analyse in this appendix the dependence of $f^{\rm NE}$ on the MB effects 
using a conventional diagrammatic NE approach for the interactions.

The lowest order non vanishing lesser and greater self-energies have the form of a
convolution product of the following type:
\begin{equation}
\label{eq:Sigmalessgtr}
\Sigma^\lessgtr_{\rm int}(\omega)  =  i  \int \frac{{\rm d}u}{2\pi} \
    \mathcal{B}^\lessgtr(u) \ G_C^\lessgtr(\omega - u) \ ,
\end{equation} 
where $\mathcal{B}(\omega)$ is related to a boson propagator.

For electron-phonon interaction, $\mathcal{B}(\omega)$ is given by
$\mathcal{B}(\omega)=\gamma_0^2 D(\omega)$, where $D(\omega)$ is the
phonon propagator and $\gamma_0$ is the electron-phonon coupling
constant.
Different levels of approximation can be used by considering the
bare phonon propagator $D_0(\omega)$, or a partially dressed phonon
propagator $\mathcal{D}_0(\omega)$ or the fully dressed phonon propagator $\mathcal{D}(\omega)$.

For electron-electron interaction, $\mathcal{B}(\omega)$ is the screened Coulomb
interaction $W(\omega)$ in which the screening is obtained according to different
levels of approximation. We describe a few of them in the following.

\emph{Electron-phonon interaction.---}
When dealing with the bare phonon, the lesser interaction self-energy becomes
\begin{equation}
\label{eq:SigmalessgtrD0}
\begin{split}
\Sigma^<_{\rm int}(\omega)  = &   i \gamma_0^2 \int \frac{{\rm d}u}{2\pi} \
    D_0^<(\omega-u) \ G_C^<(u) \\
= & - \gamma_0^2 \int {{\rm d}u} \
    D_0^<(\omega-u) \ f^{\rm NE}(u)\ A_C(u) \ .
\end{split}
\end{equation} 
Clearly such a self-energy is a functional of the NE distribution $f^{\rm NE}(u)$ 
and of the spectral function $A(u)$. One obtains a similar results for the greater
self-energy $\Sigma^>_{\rm int}(\omega)$.

For the partially dressed $\mathcal{D}_0$ or the fully dressed phonon 
propagator $\mathcal{D}$, we have the following expressions for the propagator
$\mathcal{D}_0(\omega) = D_0(\omega) + D_0(\omega) \gamma_0^2  P(\omega) D_0(\omega)$
or $\mathcal{D}(\omega) = [D_0(\omega)^{-1} - \gamma_0^2 P(\omega)]$ with $P(\omega)$
being the polarisation function.
At the lowest order, the polarisation is given by the electron-hole bubble diagram and
its lesser and greater components are
\begin{equation}
\label{eq:polarP}
\begin{split}
P^\lessgtr(\omega) = -i \int\frac{{\rm d}u}{2\pi} \ G_C^\lessgtr(u)\ & G_C^\gtrless(u-\omega) \\
= -i2\pi  \int {{\rm d}u} \ f^{\rm NE}(u) (1 - & f^{\rm NE}(u-\omega))\ A_C(u) A_C(u-\omega) \ ,
\end{split}
\end{equation} 
which is again a functional of $f^{\rm NE}$ and $A_C$. Therefore we find that for any
phonon propagator, we have
$\Sigma^<_{\rm int}=\Sigma^<_{\rm int}[f^{\rm NE},A_C]$.

\emph{Electron-electron interaction.---}
The screened Coulomb interaction $W(\omega)=v_q / \epsilon^{-1}(\omega,q)$ can be calculated within different
level of approximation for dielectric function $\epsilon^{-1}(\omega,q)$ ($v_q$ is the Fourier $q$-component of the
bare Coulomb interaction).

In the plasmon-pole approximation \cite{Hedin:1969,Ness:2011b}, we have 
$\epsilon^{-1}(\omega,q)=1+\omega_p^2/(\omega^2-\omega_q^2)$, where
$\omega_p$ is the bulk plasmon energy and $\omega_q$ the plasmon dispersion relation.
The dynamic part of the screened Coulomb potential $W(\omega)-v$ 
can be rewritten as
\begin{equation}
v_q \left( \epsilon^{-1}(\omega,q) -1 \right)
= \frac{v_q \omega_p^2}{2\omega_q}\ \frac{2\omega_q}{\omega^2-\omega_q^2}
= \gamma_p^2\ B_p(\omega,q),
\label{eq:boson_propagator}
\end{equation}
which involves a coupling constant $\gamma_p$ and the bosonic propagator
$B_p(\omega)$ of the plasmon modes.
This corresponds to the similar case of the bare phonon propagator described above.
Using the same reasoning
we find that the interacting self-energy $\Sigma^<_{\rm int}$ is a functional of
$f^{\rm NE}$ and $A_C$.

Within the $GW$ approximation \cite{Hedin:1969,Thygesen:2007,Darancet:2007,Thygesen:2008a,Rangel:2011,Ness:2011b}, 
the screened Coulomb interaction is given
by $W(\omega)=v+vP(\omega)W(\omega)$. This expression is a formally equivalent to the case of the fully 
dressed phonon propagator since 
$\mathcal{D}(\omega) = [D_0(\omega)^{-1} - \gamma_0^2 P(\omega)]=
D_0(\omega) + D_0(\omega) \gamma_0^2  P(\omega) \mathcal{D}(\omega)$.
Hence applying the previous analysis, we find again that 
$\Sigma^<_{\rm int}=\Sigma^<_{\rm int}[f^{\rm NE},A_C]$.

\emph{Vertex corrections and higher order diagrams.---}
We can also consider higher order diagrams for the electron-phonon and electron-electron, as well as 
vertex corrections to build more elaborate self-energies.
From our earlier work [\onlinecite{Dash:2010,Dash:2011,Ness:2011b}], it can been seen from the 
expressions of the second order and vertex correction diagrams that the self-energies 
$\Sigma^\lessgtr_{\rm int}$ can always be expressed as functional of the NE distribution $f^{\rm NE}(\omega)$
and of the spectral function $A_C(\omega)$ \cite{Note:6}.

\section{Lowest order expansion for the current}
\label{app:LOE}

For the two-terminal quantum devices we consider, the current $I(V)$ is given
by the famous Meir and Wingreen expression \cite{Meir:1992}:
\begin{equation}
\label{eq:I_MeirWingreen}
  \begin{split}
  I = \frac{ie}{h} \int {\rm d}\omega\  &
    {\rm Tr} 
    \left[ 
      \left( f_L(\omega) \Gamma_L - f_R(\omega) \Gamma_R \right)
      \left( G_C^r(\omega) - G_C^a(\omega) \right) 
    \right.
\\
       & \left. +
      \left( \Gamma_L(\omega) - \Gamma_R(\omega) \right) G_C^<(\omega) \right]  ,
  \end{split}
\end{equation}
where we recall that $\Gamma_\alpha(\omega)$ is the spectral function of the lead
$\alpha$ self-energy, i.e. $\Gamma_\alpha=i(\Sigma_\alpha^r-\Sigma_\alpha^a)=i(\Sigma_\alpha^>-\Sigma_\alpha^<)$.

For the single impurity model, the trace drops off and one deals with functions only.
Using the definitions $2\pi A_C=i(G_C^r-G_C^a)$ and $G_C^<=-f^{\rm NE}(G_C^r-G_C^a)$,
we obtain 
\begin{equation}
\label{eq:I_MeirWingreen_A}
  \begin{split}
  I = \frac{2\pi e}{h} \int {\rm d}\omega\ 
      ( f_L \Gamma_L - f_R \Gamma_R ) A_C - ( \Gamma_L - \Gamma_R ) f^{\rm NE} A_C .
  \end{split}
\end{equation}

The lowest order expansion of the current, in terms of elastic and inelastic processes,
is obtained by introducing the approximated form Eq.~(\ref{eq:fNE_LOE2}) for the 
NE distribution. 
The current is built on two contribution $I=I_{\rm el}+I_{\rm inel}$ with
\begin{equation}
\label{eq:I_elas}
  \begin{split}
  I_{\rm el} =&  \frac{2\pi e}{h} \int {\rm d}\omega\ 
      ( f_L \Gamma_L - f_R \Gamma_R ) A_C - ( \Gamma_L - \Gamma_R ) f_0^{\rm NE} A_C \\
=& \frac{ e}{h} \int {\rm d}\omega\ (f_L - f_R) \frac{2\Gamma_L\Gamma_R}{\Gamma_{L+R} } 2\pi A_C ,
  \end{split}
\end{equation}
the second line is simply obtained from the definition of the non-interaction NE distribution Eq.~(\ref{eq:f0NE}).
We can identify $I_{\rm el}$ in Eq.~(\ref{eq:I_elas}) as a Landauer-like current expression with the transmission 
given by the usual formula $T(\omega) = {\rm Tr}[\Gamma_L(\omega) G_C^r(\omega) \Gamma_R(\omega) G^a(\omega)]
+[\Gamma_R(\omega) G_C^r(\omega) \Gamma_L(\omega) G^a(\omega)] \equiv 
{2\Gamma_L\Gamma_R}/{\Gamma_{L+R} }\ 2\pi A_C(\omega)$.
This is a purely elastic transmission when the GFs or $A_C(\omega)$ are calculated in the absence of
interaction. In the presence of interaction, we are dealing with elastic transport with renormalised GFs \cite{Ness:2010}.

The second contribution to the current is given by
\begin{equation}
\label{eq:I_inel}
  \begin{split}
  I_{\rm inel} =  - & \frac{e}{h} (2\pi\gamma_0)^2 \int {\rm d}\omega\ \frac{\Gamma_L - \Gamma_R}{\Gamma_{L+R}} \\
  & \left[ A_C(\omega+\omega_0) f^{\rm NE}_0(\omega+\omega_0)\ A_C(\omega) [1-f^{\rm NE}_0(\omega)]  \right. \\
 -& \left. A_C(\omega) f^{\rm NE}_0(\omega)\ A_C(\omega-\omega_0) [1-f^{\rm NE}_0(\omega-\omega_0)] \ \right].
  \end{split}
\end{equation}
This is simply the lowest order inelastic contribution to the current, corresponding to vibron emission
by electron and hole.
When Eq.~(\ref{eq:I_inel}) is recast in terms of the Fermi distributions $f_L$ and $f_R$ entering
the definition of $f_0^{\rm NE}$, one recovers the lowest order expansion results obtained from
scattering theory \cite{Montgomery:2003b,Montgomery:2003a,Kim:2013} and from NEGF
\cite{Paulsson:2005,Viljas:2005,Frederiksen:2007,Rossen:2013} if the spectral function $A_C$ 
is calculated in the absence of interaction.

The important point here is that our results are obtained in a rather straightforward manner 
by using the concept of NE distribution. They are equivalent to others when working within with same
degree of approximation for the interaction self-energy. However, with the use of approximated forms
for the NE distribution, we can still perform self-consistent calculations, which go beyond 
second order perturbation theory.

\section{A two-level model}
\label{app:2levelmodel}

In this appendix, we provide an example for a model of the central region
consisting of two levels $i,j=1,2$ with two different kind of e-ph coupling, a local 
Holstein-like coupling on each site and an off-diagonal Su-Schrieffer-Heeger-like
coupling between the two levels.

The interaction self-energies are non-diagonal 2x2 matrix with elements 
$\Sigma^\lessgtr_{ij}(\omega)$
given (in the limit of low temperature) by:
\begin{equation}
\label{eq:SE2x2}
  \begin{split}
\Sigma^\lessgtr_{{\rm int},11}(\omega) & = \gamma_{0,1}^2 G^\lessgtr_{C,11}(\omega \mp \omega_1) \\
\Sigma^\lessgtr_{{\rm int},12}(\omega) & = \gamma_{0,{\rm od}}^2 G^\lessgtr_{C,12}(\omega \mp \omega_{\rm od}) \\
\Sigma^\lessgtr_{{\rm int},21}(\omega) & = \gamma_{0,{\rm od}}^2 G^\lessgtr_{C,21}(\omega \mp \omega_{\rm od}) \\
\Sigma^\lessgtr_{{\rm int},22}(\omega) & = \gamma_{0,2}^2 G^\lessgtr_{C,22}(\omega \mp \omega_2) \ ,
  \end{split}
\end{equation}
where $\omega_i$ and $\gamma_{0,i}$ are the energy and coupling constant for the local e-ph 
interaction on level $i=1,2$ and $\omega_{\rm od}$ and $\gamma_{0,{\rm od}}$ are the corresponding
quantities for the non-local e-ph interaction between level 1 and 2.

For simplicity we consider the coupling to the lead is diagonal, i.e.
$\Gamma_{L+R,ij}=\Gamma_i \delta_{ij}$ and 
therefore the non-interacting NE distribution matrix $\underline{f}_0^{\rm NE}$ is also diagonal, 
with elements $f_{0,i}^{\rm NE}$ given by Eq.~(\ref{eq:f0NEn}).

We focuss in the following on the LOE of $\underline{f}^{\rm NE}$. This approximation
still shows how the different components of the NE distribution matrix are obtained 
in the presence of a non-diagonal interaction self-energy.

Following the derivation given in Section \ref{sec:approxfNEcalc}, Eq.~(\ref{eq:fNEnm})
can be recast as

\begin{equation}
\label{eq:fNEnmLOE1}
\begin{split}
&\underline{f}^{\rm NE}(\omega) 
\sim \\
& \underline{G}^r_C 
\left( \underline{f}_0^{\rm NE} \underline{\Gamma}_{L+R} - i \underline{\Sigma}_{\rm int}^<
\right) \underline{\Gamma}_{L+R}^{-1}
\left( \underline{1} - \underline{\Gamma}_{L+R}^{-1} i \underline{\Sigma}^{>-<}_{\rm int} 
\right) (\underline{G}^r_C)^{-1} \\
& \sim 
\underline{G}^r_{0,C} 
\left( \underline{f}_0^{\rm NE} - i \underline{\Sigma}_{\rm int}^< \underline{\Gamma}_{L+R}^{-1}
- \underline{f}_0^{\rm NE} 
\underline{\Gamma}_{L+R}^{-1} i \underline{\Sigma}^{>-<}_{\rm int} \right) 
(\underline{G}^r_{0,C})^{-1} .
\end{split}
\end{equation}
where we kept only the lowest order terms, $\underline{\Sigma}^{>-<}_{\rm int}$ is a
contraction for $\underline{\Sigma}^{>-<}_{\rm int}=
\underline{\Sigma}^>_{\rm int}-\underline{\Sigma}^<_{\rm int}$  
and $\underline{G}^r_{0,C}$ is the non-interacting GF of the region $C$.
Such a GF is diagonal with elements ${G}^r_{0,i}(\omega)$ in the two-level representation.
Hence we obtain the following LOE for $\underline{f}^{\rm NE}$:
\begin{eqnarray}
\label{eq:fNEnmLOE2}
\underline{f}^{\rm NE}_{\rm LOE}(\omega)=
\left[
\begin{array}{cc}
F_{11}(\omega) & r(\omega) F_{11}(\omega) \\ 
F_{21}(\omega)/r(\omega) & F_{22}(\omega)
\end{array} 
\right] ,
\end{eqnarray}
where $r(\omega)$ is ratio $r=G^r_{0,1}(\omega)/G^r_{0,2}(\omega)$
and $F_{ij}$ are the matrix elements of 
\begin{equation}
\label{eq:Fnm}
\underline{F} = 
\underline{f}_0^{\rm NE} - i \underline{\Sigma}_{\rm int}^< \underline{\Gamma}_{L+R}^{-1}
- \underline{f}_0^{\rm NE} 
\underline{\Gamma}_{L+R}^{-1} i \underline{\Sigma}^{>-<}_{\rm int} \ .
\end{equation}

By the definition Eq.~(\ref{eq:GFfNEnm}), the matrix elements of $\underline{G}^\lessgtr_C$ entering the 
definition of the self-energies $\underline{\Sigma}^\lessgtr_{\rm int}$ depend 
on both the diagonal and off-diagonal elements of $\underline{f}^{\rm NE}$.
 
However, at the LOE, we substitute $\underline{f}^{\rm NE}$ with the non-interacting
distribution $\underline{f}_0^{\rm NE}$ which is diagonal.
Hence we have 
\begin{equation}
\label{eq:Glessgtr_LOE}
\begin{split}
G^<_{C,ij}(\omega) & = i 2 \pi f_{0,i}^{\rm NE}(\omega) A_{C,ij}(\omega) \\
G^>_{C,ij}(\omega) & = i 2 \pi (f_{0,i}^{\rm NE}(\omega)-1) A_{C,ij}(\omega) \ .
\end{split}
\end{equation}

After substitution into the definition of the self-energy Eq.~(\ref{eq:SE2x2}),
we obtain from Eq.~(\ref{eq:Fnm}) the following matrix
elements of $\underline{F}$:
\begin{equation}
\label{eq:fNEiiLOE}
\begin{split}
&F_{ii}(\omega) = f^{\rm NE}_{0,i}(\omega) \\ 
&+ \frac{2\pi\gamma_{0,i}^2}{\Gamma_i}\ \left(
  A_{C,ii}(\omega+\omega_i)\ f^{\rm NE}_{0,i}(\omega+\omega_i)\ [1-f^{\rm NE}_{0,i}(\omega)] \right. \\
&\left. - A_{C,ii}(\omega-\omega_i)\ [1-f^{\rm NE}_{0,i}(\omega-\omega_i)]\ f^{\rm NE}_{0,i}(\omega)\ \right) \ ,
\end{split}
\end{equation}
for the diagonal elements ($i=1,2$) and for the off-diagonal elements:
\begin{equation}
\label{eq:fNE12LOE}
\begin{split}
&F_{12}(\omega) =  \\
& \frac{2\pi\gamma_{\rm od}^2}{\Gamma_1}\ \left( 
  \left[\frac{\Gamma_1}{\Gamma_2} - f^{\rm NE}_{0,1}(\omega)\right] f^{\rm NE}_{0,1}(\omega+\omega_{\rm od})\ 
   A_{C,12}(\omega+\omega_{\rm od})\ \right. \\
&\left. - A_{C,12}(\omega-\omega_{\rm od})\ [1-f^{\rm NE}_{0,1}(\omega-\omega_{\rm od})]\ f^{\rm NE}_{0,1}(\omega)\ \right) \ .
\end{split}
\end{equation}
The matrix element $F_{21}$ is obtained from the expression of $F_{12}$ by swapping the indices $1\leftrightarrow 2$.

From Eqs.~(\ref{eq:fNEiiLOE}-\ref{eq:fNE12LOE}) and (\ref{eq:fNEnmLOE2}), 
we can see that the diagonal elements ${f}^{\rm NE}_{{\rm LOE},ii}$ are real and given by
an expression similar to the result Eq.~(\ref{eq:fNE_LOE2}) obtained for the single-level model.
The off-diagonal elements ${f}^{\rm NE}_{{\rm LOE},ij}$ acquire an imaginary part
via the presence of the ratio $r(\omega)$. 
In some cases, such an imaginary can be negligible or even vanishing.

The interesting point in the LOE is that each matrix element ${f}^{\rm NE}_{{\rm LOE},ij}$
is to be determined self-consistently with the corresponding matrix element $A_{C,ij}$ of the spectral
function. There is no mixing between the different $A_{C,ij}$ and ${f}^{\rm NE}_{{\rm LOE},ij}$.

Obviously, beyond the LOE, there will be some mixing between the different matrix elements
of the NE distribution and the spectral function, since $\underline{G}^r_C$ is generally not
diagonal and $\underline{G}^<_C$ is given by
\begin{eqnarray}
\label{eq:Glessnm}
\underline{G}^<_C = i 2 \pi 
\left[
\begin{array}{cc}
f^{\rm NE}_{11} &  f^{\rm NE}_{12} \\ 
f^{\rm NE}_{21} &  f^{\rm NE}_{22} \\ 
\end{array} 
\right]
\left[
\begin{array}{cc}
A_{C,11} &  A_{C,12} \\ 
A_{C,21} &  A_{C,22} \\ 
\end{array} 
\right] \ .
\end{eqnarray}


\begin{thebibliography}{73}
\bibitem{Note:1}
With the behaviour of the
NE distributions, we have $f^{{\rm NE}(n)}(\omega)/f^{{\rm NE}(n+1)}(\omega)\rightarrow 1$
when $\omega<\omega_{\rm min}$ and $\omega>\omega_{\rm max}]$.
\bibitem{Note:2}
This is not the case when the electron temperature $kT_{L,R}$ is much larger that 
$\omega_0$, but this is a very special transport regime we are not considering here.
\bibitem{Note:5}
Equality between the two expressions in Eq.~(\ref{eq:SigmaFock_ratio2}) would hold 
only in the limit $\omega_0\rightarrow 0$.
But this corresponds to a pathological case of phonon with zero
frequency and generates only an infinite static renormalisation, and no inelastic self-energies,
$\Sigma_{\rm int}^{\lessgtr}(\omega) \equiv 0$, which is not relevant for any study and is
far from any realistic conditions.
\end{thebibliography}
\end{document}